\documentclass{JFM-FLM_Au}
\usepackage{xcolor}

\usepackage{ulem}

\lefttitle{T. Gaichies, A. Salonen, A. Antkowiak, E. Rio}
\righttitle{Journal of Fluid Mechanics}

\title{Shape of an interface hit by an oblique jet}

\author{Theophile Gaichies\aff{1}, Anniina Salonen\aff{1,2} Arnaud Antkowiak\aff{3}\and Emmanuelle Rio\aff{1}}

\affiliation{\aff{1} Laboratoire de Physique des Solides, UMR 8502, CNRS, Université Paris-Saclay, 91405 Orsay, France
\aff{2}Soft Matter Sciences and Engineering, ESPCI Paris, PSL University, CNRS, Sorbonne Université, 75005 Paris, France
\aff{3} Institut Jean le Rond $\partial$'Alembert, Sorbonne Université, CNRS, F-75005 Paris, France}

\corresau{Theophile Gaichies, \email{theophile.gaichies@gmail.com}}

\begin{document}
\maketitle

\begin{abstract}

We report on the shape taken by the interface of a liquid bath when hit by a smooth oblique steady jet. When the angle between the jet and the bath decreases below $50^\circ$, a cavity is formed in front of the jet. In the inertial regime we explore, the jet boundary layer detaches in the impact region, thereby delimiting a core jet region outside of which the liquid is mainly in hydrostatic equilibrium. The shape of the outer meniscus is shown to be related to the one outside a tilted fiber piercing the fluid interface. In order to unravel the flow features and separation, we perform direct numerical simulations  and show that the flow detachment displays an asymmetry, which results in the acceleration of the liquid below the surface, thereby creating a depression. With this observation, we propose a model balancing the suction force of this depression with the weight of the displaced water and the surface tension force to obtain a prediction for the typical width of the cavity. 
\end{abstract}

\begin{keywords}
Authors should not enter keywords on the manuscript, as these must be chosen by the author during the online submission process and will then be added during the typesetting process (see \href{https://www.cambridge.org/core/journals/journal-of-fluid-mechanics/information/list-of-keywords}{Keyword PDF} for the full list).  Other classifications will be added at the same time.
\end{keywords}


\section{Introduction}

Air entrainment processes play a critical role in various natural and industrial contexts, in which a liquid jet impacts a liquid surface (\cite{kiger2012air,deike2022mass}). 
In such systems, characterizing the shape of the interface is essential for predicting entrainment behavior. 
A classic example is the vertical viscous jet. In this configuration, viscous stresses induce a cusp whose curvature is governed by the capillary number (\cite{jeong1992free}). 
By analyzing the airflow within this cusp, \cite{eggers2001air} successfully predicted the threshold velocity for air entrainment, a result subsequently validated by experimental studies (\cite{lorenceau2003fracture, lorenceau2004air}). 
Furthermore, when the jet speed exceeds the threshold velocity, a symmetric air sheet is entrained by the jet and destabilises in a cloud of bubbles. By studying the shape of the interface, \cite{lorenceau2004air} showed that this system was analogous to the Landau Levich Derjaguin problem (\cite{derjaguin1943thickness,levich1942dragging}), 
and were then able to predict the thickness of the air sheath and thus the total flux of entrained air.

In contrast, for low-viscosity liquids such as water, the free-surface dynamics of plunging jets remain less well understood, complicating the prediction of entrainment thresholds and fluxes \citep{kiger2012air,sene1988air,Antkowiak2021}. The interface below the jet consists of a bubble cloud, whose depth can be predicted through a momentum flux balance  \citep{clanet1997depth}, taking into account the void fraction of the bubble cloud \citep{dev2024liquid,dev2025experimental,guyot2020penetration}. The jet's interface often becomes non-stationary due to perturbations propagating along it, which govern the entrainment mechanism \citep{zhu2000mechanism,redor2025air}. When these perturbations are suppressed in millimetric vertical jets, only an upward meniscus is typically observed, without any air entrainment \citep{patrascu}. 

However, when axisymmetry is broken, either through jet translation \citep{chirichella2002incipient} or inclination \citep{koga1982bubble,miwa2019experimental,detsch1990critical,deshpande2012computational}, a cavity can form and favor air entrainment. While such cavities have been documented in the literature, there remains a significant gap in quantitative modelling to describe such interfacial shapes.

In this article, we choose to introduce an asymmetry by inclining the plunging jet and we focus on the shape of the interface resulting from this impact.  We first examine a simpler case, where the jet is replaced by a glass fiber, to investigate the effect of the introduced asymmetry on the meniscus' shape. We then turn to direct numerical simulations of the flow to obtain the velocity field in the region connecting the jet to the interface. We observe that the flow separation from the interface is asymmetric, resulting in an acceleration of the liquid and a depression beneath the interface. With this information, we propose a simple model to capture the dependence of the cavity's width on the system parameters.  

\section{Experimental observation of the cavity}
\subsection{Experimental setup}
The liquid jet used throughout this study is typically set into motion by an overpressure applied by a pressure controller (Elveflow OB1 Mk4) on a water volume contained in a glass bottle (Schott). This forcing drives the liquid through the tubing connected to a stainless steel nozzle (Nordson EFD Optimum). The inclination of the nozzle is adjusted to produce a jet impacting the water bath at an angle $\alpha \in [21,49]^\circ $. By changing the nozzle, the radius of the jet $R$ is varied between 0.12 mm and 0.73 mm. The flow rate is measured using a graduated cylinder and a timer, allowing the determination of the mean speed of the jet $V \in [0.8; 5.4]\ \rm m.s^{-1}$.

The bath is contained in a cubic quartz tank (6 cm side length). To ensure repeatability of the experiments, the water must be free of contaminants. For this reason, all tubing used in the setup is made of fluoropolymer (Versilon FEP, Saint Gobain), the water is ultrapure (resistivity greater than 18.2 $\rm M\Omega.cm$) and the bath is changed at least every thirty minutes. 

We observe the interface using a LED backlight (PHLOX) and a digital camera (Basler a2A1920-160umPRO) equipped with a lens (telecentric SilverTL 0.75x when the jet radius $R$ is less than 0.25 mm, and Ricoh Lens FL-CC5028-2M 50 mm with an 8 mm extender ring otherwise). To be able to image the interface without deflection by menisci on the glass, we only keep the images taken when the tank is filled to the brim with water so that the interface is nearly flat. 
\label{exp_meth}

\subsection{Observation of the cavity}

\begin{figure}[h!]
    \centering
    \includegraphics[width=\textwidth]{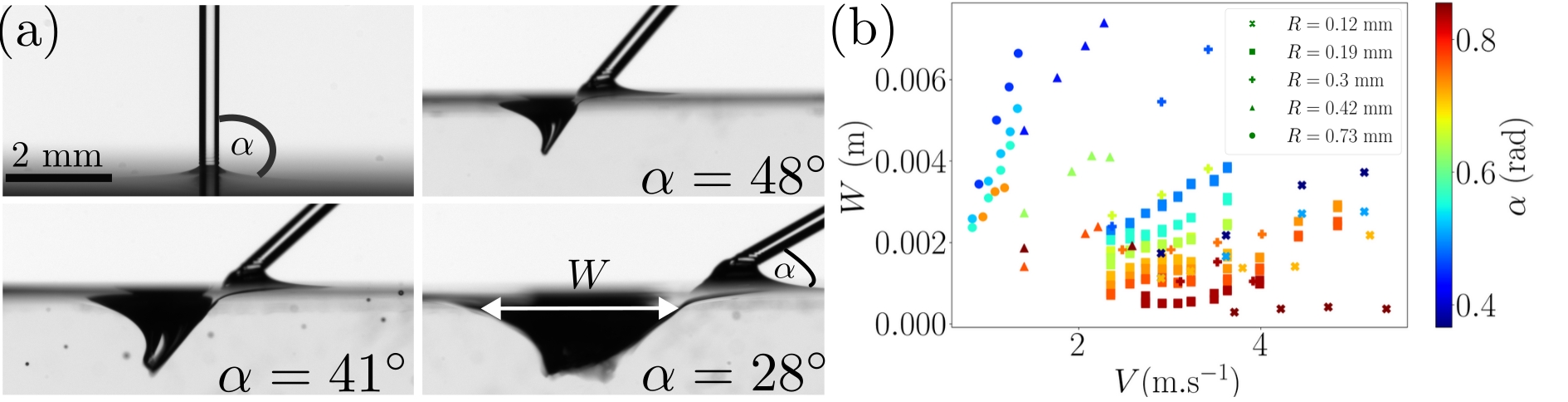}
    \caption{(a) Experimental images of the interface impacted by a jet of radius $R = 0.19$ mm with a speed $ V = 2.8\ \mathrm{m.s^{-1}}$, with various angles $\alpha$ between the jet and the bath. (b) Measurements of the width of the cavity $W$ for jets of varying speeds, radii, and inclination with respect to the bath.}
    \label{fig:intro_jet}
\end{figure}
As shown in figure \ref{fig:intro_jet}(a), when the jet impacts the bath vertically, we observe an axisymmetric meniscus in all our experiments. This is consistent with observations of smooth millimetric jets in the literature (\cite{patrascu,hancock2002fluid,gaichies2024effective}). In this case, the flow separates from the interface, and the outer part of the interface can be described by hydrostatic equations \citep{gaichies2024effective}. 

However, when the jet inclination $\alpha$ reaches around $50 ^\circ$, a cavity appears in front of the jet. As the jet approaches the horizontal, the cavity widens. To characterize this evolution, we define the typical width of the cavity $W$, measured at a depth below the interface equal to 10\% of the distance between the interface and the deepest point of the cavity. Due to the waves traveling along the cavity, the typical uncertainty on $W$ can be up to 10\% for large cavities. The experimental measurements are displayed in figure~\ref{fig:intro_jet}(b) for various radii, speeds, and inclinations. The cavity is wider for larger and faster jets. Crucially, the experimental data highlight that the jet inclination is the dominant parameter governing the appearance and geometry of the cavity. The  cavity width is up to 10 times larger for very inclined jets. 

We will see in section \ref{sec:num} that the flow separation from the interface is still present in this inclined configuration. We can then expect that, similarly to the case of the vertical jet, parts of the interface can be described by hydrostatic equations. It is thus relevant to study the static equivalent of our system, an inclined fiber in a liquid bath, to isolate the effect of inclination on the shape of a meniscus.

\section{Asymmetric meniscus on a fiber}
\label{sec:AsymmetricMeniscus}

\begin{figure}[h!]
    \centering
    \includegraphics[width=0.99\textwidth]{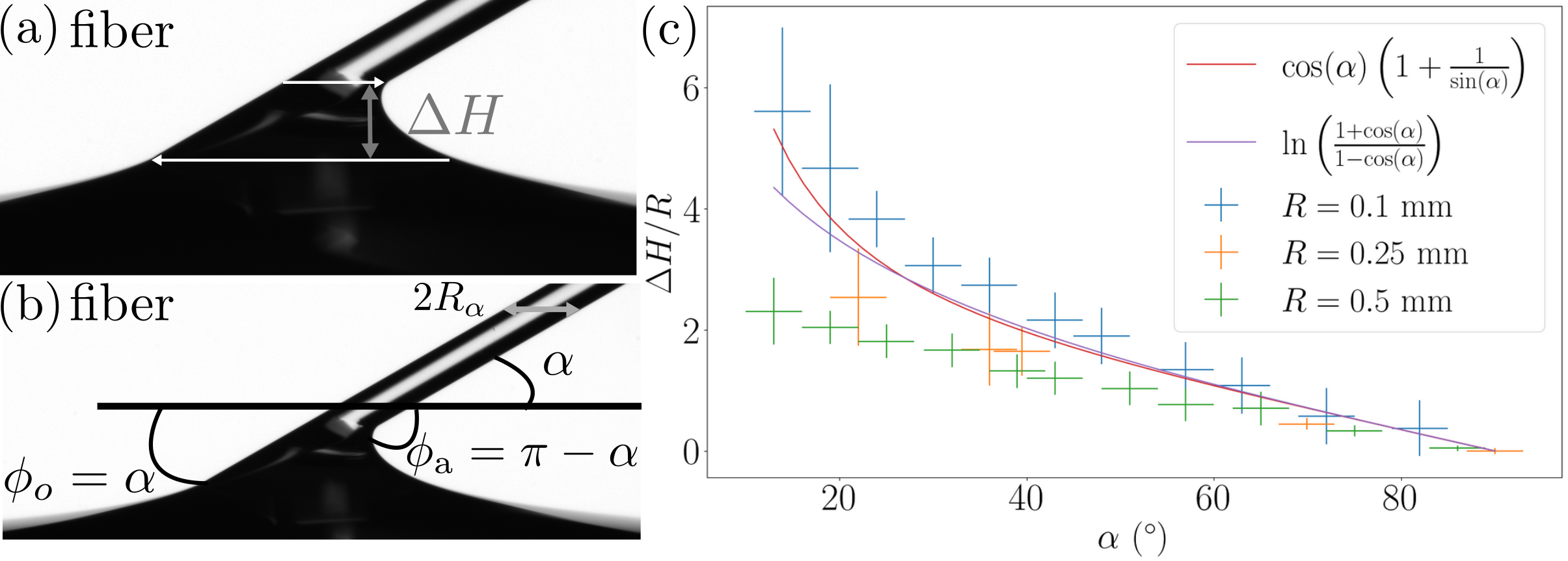}
    \caption{(a) Experimental image of the meniscus on a glass fiber of radius $R = 0.1$ mm and inclination $\alpha = 30^\circ$ in a pool of silicone oil. The two white arrows designate the point where the meniscus meets the fiber. (b) Sketch representing the angles $\phi_0$ and $\phi_a $ between the fiber and the horizontal. (c) Measurements of the maximum height difference normalized by the fiber radius $\Delta H /R$. The red curve corresponds to the empirical model of \cite{raufaste2013deformation}, and the blue line represent equation \ref{eqmodjames}.}
    \label{fig:fiber}
\end{figure}

In this section, we study the shape of the meniscus on an inclined fiber. The jet is replaced by a glass fiber (Hilgenberg) of radius $R \in [0.1, 0.25, 0.5]$ mm and the water bath is replaced by a silicone bath of similar viscosity (Carl Roth  M 3 cSt) to achieve a near zero contact angle with the fiber. When the fiber is inclined, an asymmetric meniscus is observed (figure \ref{fig:fiber}(a)), without any cavity. The meniscus height varies along the fiber circumference, and is greater on the side with the acute angle than the obtuse one. The maximum height difference $\Delta H$ between these two points on the contact line can be measured, as indicated on the figure and plotted in figure \ref{fig:fiber}. This asymmetry decreases with the angle $\alpha$ reaching a uniform height for a vertical fiber. To model the evolution of $\Delta H$ with $R$ and $\alpha$, we use the formula for the meniscus' height on a vertical fiber given by  \cite{james1974meniscus}: 
\begin{equation}
    H(\phi) \underset{\varepsilon \to 0}{\sim} R \sin(\phi)\left[\ln\left(\frac{4}{\varepsilon(1+\cos(\phi))}\right) -\Gamma\right],
\end{equation}

\noindent where $\Gamma$ is Euler's constant, $\varepsilon = (\rho gR^2 /\gamma)^{1/2}$ (where $g$ is the acceleration due to the gravity, $\rho$ the density of the liquid and $\gamma$ its surface tension) is the square root of Bond number and $\phi$ is the angle of the liquid on the fiber with respect to the horizontal. In the model developed by James, the presence of an angle is due to a non-zero contact angle $\theta$ and $\phi =\pi/2 -\theta $. 

To obtain a prediction for $\Delta H$, we propose to adapt this formula to our problem, with the assumption that it will hold locally despite the loss of axisymmetry. In this inclined configuration, the angle between the interface and the horizontal used in James' model is fixed by the fiber inclination rather than by the contact angle. Assuming a zero contact angle we then have different $\phi$ on the two sides of the fiber: $\phi_0 = \alpha$ in the obtuse region, and $\phi_a = \pi - \alpha$ in the acute part. We also take into account that in the horizontal plane, the fiber has an apparent major axis $2R_\alpha =2R/\sin(\alpha)$. With these two ingredients, we obtain $\Delta H = H(\pi - \alpha) - H(\alpha)$ and after simplifications:
\begin{equation}
    \Delta H \approx   R\ln \left(\frac{1+\cos(\alpha)}{1-\cos(\alpha)}\right). 
\label{eqmodjames}
\end{equation}
This prediction is plotted against the experimental measurements in figure \ref{fig:fiber}(c). The agreement is satisfactory for the smallest fiber ($R = 0.1$ mm) but overestimates $\Delta H$ for larger fibers. A reasonable explanation for this discrepancy is that James' formula is only valid for vanishing $\varepsilon$. Our prediction can also be compared with the empirical formula of \cite{raufaste2013deformation}, obtained with simulations of liquid films pierced by inclined fibers, which, using our notation, is $\Delta H = R \cos(\alpha)(1+1/\sin(\alpha))$. We can see in figure \ref{fig:fiber}(c) that it agrees very well with our model down to $\alpha = 25^\circ$. This deviation at large inclination is probably due to our assumption of local axisymmetry which becomes less valid as $\alpha$ decreases. 

Having demonstrated that the inclination results in an asymmetry in the attachment height of the meniscus, we now return to jet impacts. 
Our hypothesis is that the asymmetry of the interface, with a meniscus in the acute part and a cavity in the obtuse part, is reminiscent of the asymmetry observed in the static situation.
However, to understand the coupling between the asymmetric interface and the flow we need to investigate the velocity field. 
Since, as can be seen on figure \ref{fig:intro_jet}(a), the connection between the jet and the meniscus is curved and thus hinders particle imaging velocimetry, we perform numerical simulations of the flow to access this field.
\section{Velocity field in the meniscus}
\label{sec:num}

To obtain the velocity field in the region where the curvature prevents particle tracking, we perform direct numerical simulation of the flow, using the open source software Basilisk (\cite{popinet2009accurate}). This solver uses a volume of fluid method with sharp interface reconstruction (\cite{popinet2018numerical}). It has been extensively validated, especially in configurations involving violent events, e.g. drop impact (\cite{wang2023analysis,Howland2016}), or breaking waves and bubble entrainment (\cite{Mostert2022}).
We use the momentum conserving variant of the Navier-Stokes solver. The speed of the jet at injection and the radius of the jet are the units for speed and distance. 
Instead of inclining the jet, we incline the interface by changing the orientation of the gravity vector $\vec{g}$. 
Depending on $\alpha$, we have to impose a contact angle different from $90^\circ$ on some sides of the boxes using height functions, see e.g. \cite{afkhami2008height}.
The simulations are performed in 3D using an adaptive octree grid with a refinement criterion based on the velocity and fraction field errors. 
The size of the simulation box is varied between 30 and 50 radii as the cavity gets bigger to allow the interface to relax, while optimizing the resolution of the simulation.  
As the simulations are computationally demanding, the box was only divided up to 512 times in each direction.  
This resolution ensures that the jet radius is resolved by at least 10 cells, which is sufficient to capture the topology of the flow. 
We thus focus our analysis on the structural features of the velocity field and the resulting pressure distribution mechanisms.

The simulations are performed for a Reynolds number $\rm Re = \rho V R/\eta = 400$ (with $\eta$ the dynamic viscosity of the liquid), a Weber number $\rm We = \rho V^2 R/\gamma= 12$, and a Froude number $\rm Fr = U/\sqrt{gR} =40$. These numbers correspond to a radius $R = 0.22$ mm and a speed $V = 1.85 \ \rm m.s^{-1}$, typical of the values encountered in the experiments, for a liquid with $\eta = 1\ \rm mPa.s$, $\rho = 1000 \rm \ kg.m^{-3}$ and $\gamma = 60\rm \ mN.m^{-1}$, similar to the properties of water. 

\begin{figure}[h!]
    \centering
    \includegraphics[width=0.95\textwidth]{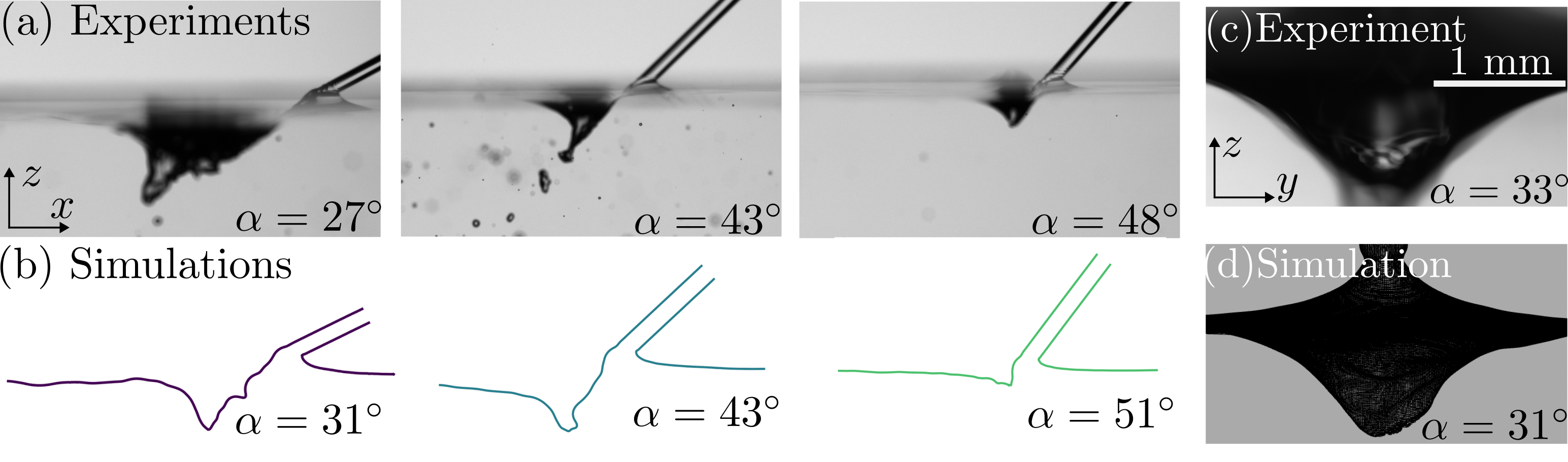}
    \caption{(a) Experimental image of a $R = 0.3$ mm jet impacting a water bath at a speed $V = 3.1\ \rm m.s^{-1}$ for various $\alpha$. (b) Interfacial profile in the x-z plane (at $z=0$) extracted from simulations of a jet with Re = 400, We = 12 and Fr = 40, for various $\alpha $. (c) Image of the cavity formed by a $R = 0. 19\rm\ mm$ jet impacting a water bath at $V = 3.2 \rm \ m.s^{-1}$ with an angle $\alpha = 33^\circ$, in the yz plane facing the impacting jet. (d) Snapshot of the reconstructed interface from the simulations with $\alpha = 31^\circ$.}
    \label{fig:comparaison_simu}
\end{figure}

In figure \ref{fig:comparaison_simu}(a) and (b), experimental images and simulated interfaces are compared, in the x-z plane (which is the symmetry plane of the system).  The same qualitative behavior is observed, with cavities widening as $\alpha$ decreases.
The shape in the zy plane (figure \ref{fig:comparaison_simu}(c) and (d)) also matches that observed experimentally.

\begin{figure}[h!]
    \centering
    \includegraphics[width=\textwidth]{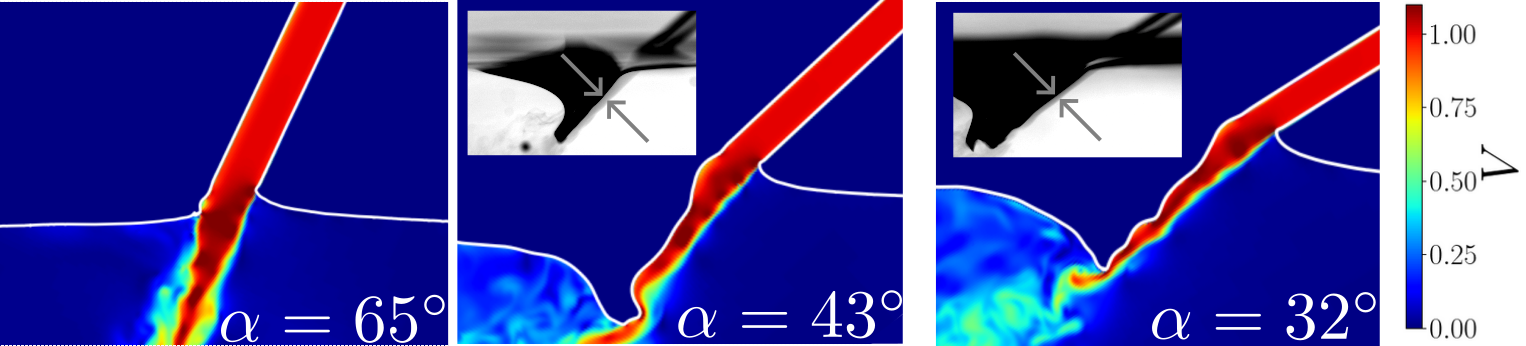}
    \caption{Simulated velocity fields in the x-z plane for various inclinations $\alpha$. The inserts are experimental images of a dyed (0.5 wt$\%$ indigo carmine) jet ($R = 0.19$ mm, $V = 2.8\mathrm{m.s^{-1}}$) impacting the interface, where the thickness of the dyed layer is highlighted with arrows.}
    \label{fig:vitesse}
\end{figure}

Pressure and velocity measurements are readily accessible in the simulations. Figure~\ref{fig:vitesse} presents a cross-section of the velocity norm in the symmetry plane. 
For $\alpha = 65^\circ$, there is an asymmetric meniscus around the jet, akin to the one on the inclined fiber in section \ref{sec:AsymmetricMeniscus}.
This validates our hypothesis that the glass fiber is a useful experiment to explore the impact of the inclination on the interface.
We can see that the velocity fields detach from the interface where the jet connects to the meniscus, as is observed for similar vertical jets (\cite{sebilleau2009flow,gaichies2024effective}). 
However, with an inclined jet, this detachment is asymmetric, and the flow separates from the interface higher in the acute part than in the obtuse part. 
Because of this asymmetry, the streamlines get closer after the detachment from the acute part. 
This leads to velocities higher than the injection speed below the separation. 
As $\alpha$ is reduced, this contraction is more pronounced, as can be seen in the simulations, where the thickness of the high velocity jet in the x-z plane gets smaller. 

We also checked this observation experimentally, by dying the solution in the pressure reservoir with 0.5 wt$\%$ indigo carmine to produce a stained jet. 
As can be seen in the insets of figure \ref{fig:vitesse}, the thickness of the colored band is thinner than the diameter of the jet, and thins as the jet is inclined, similarly to what is observed numerically. 
The qualitative agreement between experiments and simulations encourages the use of simulations for a better understanding of the experiments. 
We note that these experiments give us no quantitative information on the velocity, since the observation is only 2D and the flow is actually three-dimensional.
We will see in figure \ref{fig:PV} that the section of the jet deforms when it enters the bath leading to a flattened jet.

Another feature of this flow is that velocities near the interface are very low in the regions detached from the velocity field. In the acute part the velocities are 40 times lower than the injection speed, similar to what is observed in menisci connected to vertical jets (\cite{gaichies2024effective}), while they are five times lower than injection speed in the obtuse part. 

From the simulations, we draw two main conclusions. 
First, the velocity field separates from the interface asymmetrically, leading to an acceleration. This asymmetry is reminiscent of the asymmetry of the meniscus on an inclined fiber identified in section \ref{sec:AsymmetricMeniscus}. 
Second,  the velocity is very low close to the interface in the outer part of the flow.
In the next section, we will use these two observations to propose analytical models for the observed interfacial shape. 

\section{Modeling the shape of the interface}
\subsection{Model for the outer part of the interface}
\begin{figure}[h!]
    \centering
    \includegraphics[width=\textwidth]{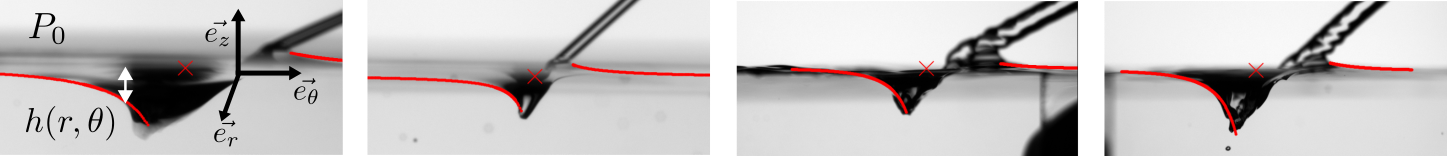}
    \caption{Images of the interfaces, with the adjusted equation \ref{a1eq} represented in red. The origin for the fit is marked with a red cross. The conditions for the jets are, from left to right: ($R = 0.12\rm\ mm$, $V = 3.6 \rm \ m.s^{-1}$, $ \alpha = 21^\circ$), ($R = 0.12\rm\ mm$, $V = 2.9 \rm \ m.s^{-1}$, $ \alpha = 41^\circ$), ($R = 0.73\rm \ mm$, $V = 0.9 \rm \ m.s^{-1}$, $ \alpha = 26^\circ$), ($R = 0.73\rm\ mm$, $V = 1.2 \rm \ m.s^{-1}$, $ \alpha = 42^\circ$).}
    \label{fig:A1}
\end{figure}
As the liquid velocity in the outer part, which is separated from the inner, fast-flowing region, is low, we propose to describe this part of the interface with hydrostatic equations.
With the notations of figure \ref{fig:A1}, we write the projection along the $z$ axis of the equilibrium between the Laplace pressure at the interface $P_0+\gamma \kappa$, with $\kappa$ the curvature and the hydrostatic pressure $P_0 -\rho g z$ to obtain, at $z=h(r,\theta)$ :
$$ \rho gh = -\gamma \kappa.$$ Within the slender slope approximation we write $\kappa = -\Delta h(r,\theta)$. We use $l_c=(\gamma/\rho g)^{1/2}$ as a characteristic length and define the nondimensional height as $\bar h= h/l_c$. Our equation can then be rewritten as the non-dimensional Poisson equation $\Delta \bar h=\bar h$. Assuming azimuthal symmetry of the form $\bar h = H(r)\exp(im\theta)$, we obtain a Bessel equation for $H(r)$:
\begin{equation}
    \bar{r}^2H'' + \bar r H' - (m^2 + \bar r^2)H = 0.
\end{equation}
The solutions of these equations are the modified Bessel functions $I_m$ and $K_m$. As the $I_m$ diverge for $\bar r \to \infty$, this leaves $K_m$ as the only candidates fulfilling the boundary conditions. The symmetry of our system, which goes from a meniscus in the acute part ($\theta = 0$), to a cavity in the obtuse part ($\theta = \pi$), suggests $m=1$. As a result, the final dimensionalized solution takes the form: 
\begin{equation}
    h(r,\theta) = A_1l_cK_1(\frac{r}{l_c})\Re(e^{i\theta}).
\label{a1eq}
\end{equation}
To test this model, we fit it on the profile of the interfaces extracted from various experimental images, leaving $A_1$ as a fitting parameter. The origin ($r=0$) is chosen manually, as the fitting procedure failed when it was left as a free parameter. Its position is however severely constrained as the fit is done on both the acute and obtuse part simultaneously. The origin is always situated in the middle of the straight part of the cavity, as indicated by a cross in the images of figure \ref{fig:A1}. These fits are represented on the experimental images in figure \ref{fig:A1}. We observe that the function $h(r,\theta)$ describes the outer part of the meniscus and the cavity simultaneously, with a single fitting parameter, which validates the hydrostatic modeling of the interface in the outer zone. 

\subsection{Model of the width of the cavity}
\begin{figure}[h!]
    \centering
    \includegraphics[width=\textwidth]{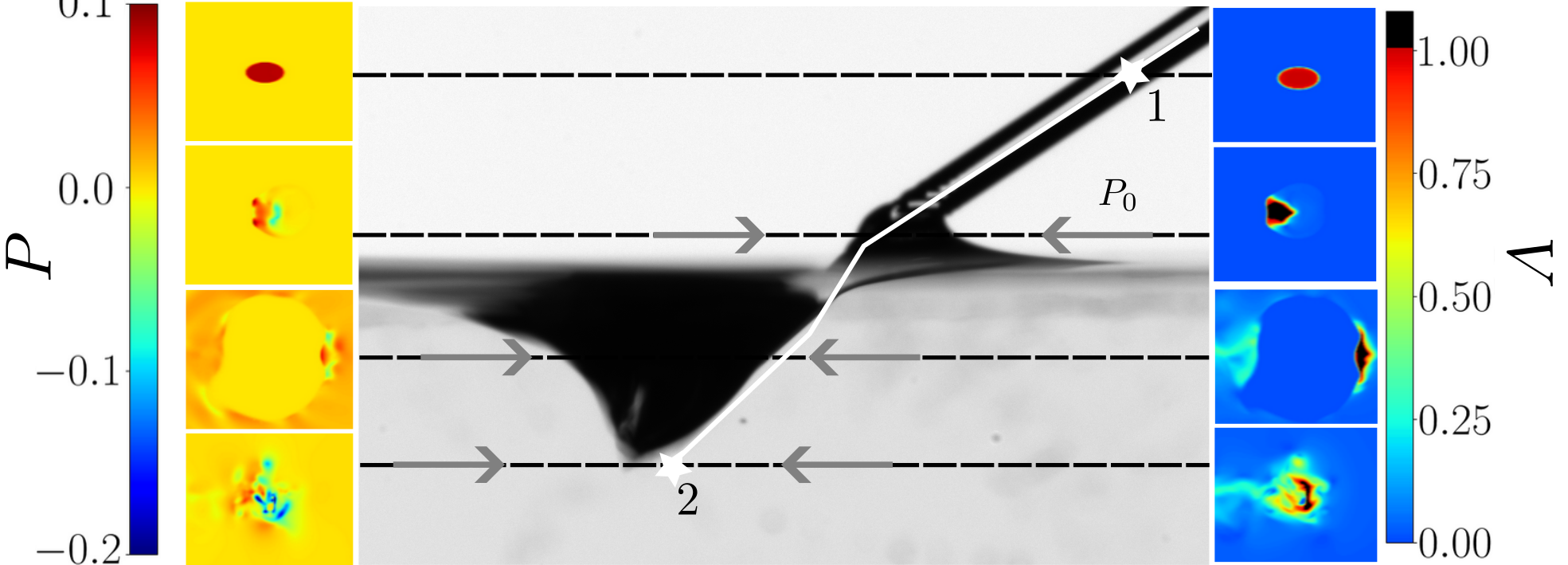}
    \caption{Pressure fields (minus the hydrostatic pressure) (on the left) and velocity fields (on the right) in the x-y plane for various height $z$, from the simulation of a jet impacting the bath with $\alpha = 31^\circ$. The increase of the velocity after impact is highlighted by the color-code, where values superior to the injection velocity are shown in black. The experimental image is of a jet ($R = 0.19\rm\ mm$, $V = 2.8 \rm \ m.s^{-1}$, $ \alpha = 33^\circ$) impacting a water bath. The white line is a sketch of the streamline used to write Bernoulli's equation. Grey arrows indicate the approximate width of the fields plotted on the side.}
    \label{fig:PV}
\end{figure}

We propose that the driving force behind the cavity's creation is the contraction of the velocity field, which lowers the pressure below the interface, creating a pulling force on the interface. 
To verify this claim, we return to the numerical simulations. We  plot in figure \ref{fig:PV} the velocity fields and the pressure deviation from hydrostatics, in the x-y plane. 
We can see that above the bath level (at point 1), the velocity of the liquid in the jet is equal to the speed of injection. The local overpressure is due to the jet's curvature.
Along the cavity, the velocity increases and is concentrated near the free surface. This is denoted by the black color in the velocity maps, black corresponding to velocities higher than the injection velocity.  
Below the cavity (at point 2), we observe a depression that correlates partly with the region of high speed. 
The simulations thus give us the hint that the depression results from the liquid acceleration, which originates from the asymmetric detachment of the velocity field from the interfaces, which is due to the asymmetric meniscus height. 
We propose that the velocity under the cavity (at point 2) $V_2$ scales with inclination in the same manner as $\Delta H$. We thus write the simplest possible scaling for the speed $V_2 \sim \beta f(\alpha) V$, with  $\beta$ a proportionality coefficient to be fitted and $$ f(\alpha) = \ln \left( \frac{1+\cos(\alpha)}{1-\cos(\alpha)}\right),$$ the geometric factor obtained in equation \ref{eqmodjames}.

If we follow the streamline represented in figure \ref{fig:PV}, we can write at point 1 Bernoulli's equation $C = P_0+ \rho V^2/2$, with $P_0$ the reference pressure at $z=0$.
We neglect the curvature of the jet, which leads to an overpressure $\gamma/R \sim 0.072/ 0.0002 \sim 360 $~Pa which is negligible in front of the typical dynamic pressure $\rho V^2/2 \sim 2$kPa.

\begin{figure}[h!]
    \centering
    \includegraphics[width=\textwidth]{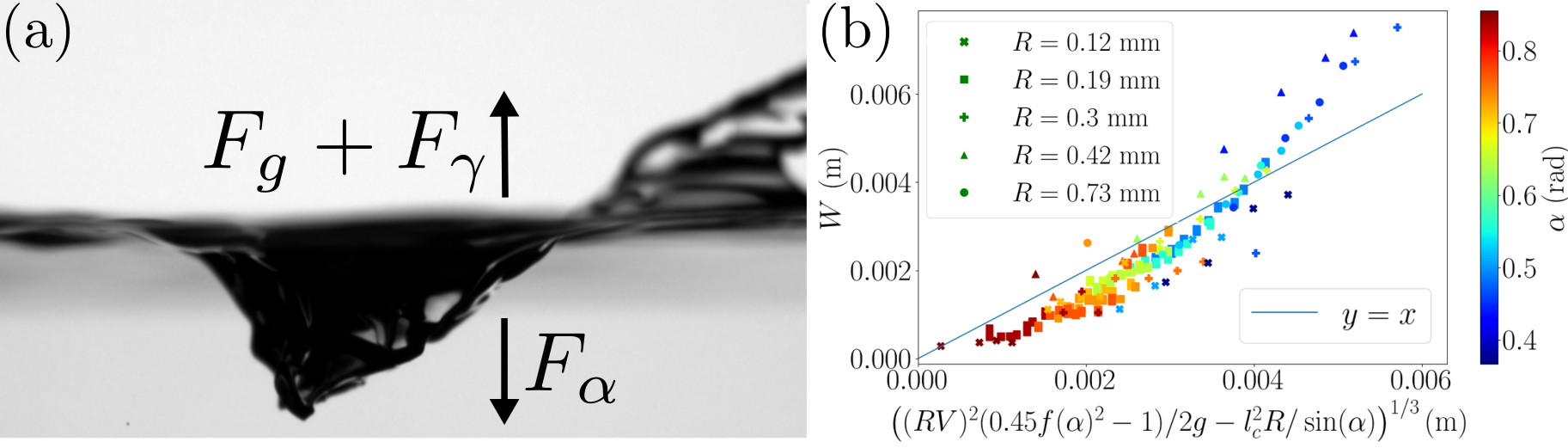}
    \caption{(a) Sketch of the force balance on the stationary cavity. (b) Comparison of the measured cavity's width against the prediction of equation \ref{eq_final_model}
    }
    \label{fig:modele}
\end{figure}

At point 2, we have $C = P_j + \rho (Vf(\alpha))^2 /2+ \rho gz$, where $P_j$ is the local pressure. 
The deviation from hydrostatic pressure under the cavity is then $P_D = P_0 -\rho g z - P_j = \frac{1}{2}\rho V^2(\beta^2 f(\alpha)^2 - 1)$. 
The suction force at the origin of the cavity creation then scales as $ R^2 P_D$. 
The opposing forces are the weight of the displaced water $F_g$ and the force exerted by surface tension $F_\gamma$. We propose to write $F_g \sim \rho g W^3$, $W$ being the typical length scale characteristic of the cavity.
To have an estimate of $F_\gamma$, we adapt the work of \cite{raufaste2012deformation}, where they find, through simulations, that a soap film exerts a purely vertical force on a tilted cylinder (of radius $R_c$) $ F_\gamma=2\gamma R_c\pi/\sin(\alpha) \sim \gamma R /\sin(\alpha)$. 
Finally, by equating the driving and opposing forces, as sketched on figure \ref{fig:modele}(a), we obtain a prediction for $W:$ \begin{equation}
    W = \left( \frac{(RV)^2 (\beta^2f(\alpha)^2 -1)}{2g} - l_c^2R/\sin(\alpha) \right)^{1/3}.
\label{eq_final_model}
\end{equation}

A comparison of the measured $W$ with the model is shown in figure \ref{fig:modele}(b), with $\beta = 0.68$ for all experiments. We can see that this simple model gives a reasonable estimate of the cavity size. However, the model overestimates the size of small cavities, and underestimate the size of cavities with $W > 4$ mm. This estimation would be improved with more delicate scalings of the cavity's volume and surface.

\section{Conclusion}
In this article, we investigated the deformation of a liquid interface impacted by an oblique jet. We showed that a cavity is created in front of the jet, while a meniscus is attached to the acute part of the jet. We studied a related system where the jet is replaced by an inclined fiber, and observed that in this case there is no cavity but an asymmetric meniscus. With a model adapted from the equation of a meniscus reaching a fiber with a finite contact angle, we found an analytical expression of the maximum height difference as a function of the fiber radius and of the angle of the inclined fiber with the bath. 

Numerical simulations revealed two distinct features of the velocity field: (i) a low speed region in the outer part of the cavity and the meniscus, and (ii) a contraction of the streamlines because of the asymmetry of the flow separation, leading to a depression responsible for the onset of a cavity. The former justified the use of hydrostatics to describe the outer part of the interface, while the latter was the missing ingredient to write a force balance that predicts the evolution of the cavity's width. 

While the current force balance offers a robust first-order prediction, the model could be refined further. In particular, a better estimation of the cavity's volume and surface would allow a more precise modeling of $F_g$ and $F_\gamma$. However, we think that this first description of the cavity shape is needed to better understand the bubble creation mentioned in the introduction.
\bibliography{jfm}
\bibliographystyle{jfm} 

\end{document}